\documentclass[pdftex]{article}
\usepackage{icrctc07}

\title{Configuration studies for a cubic-kilometre deep-sea neutrino telescope - KM3NeT - with NESSY, a fast 
and flexible approach}
\shorttitle{Configuration studies for a cubic-kilometre neutrino telescope}
\authors{J. Carr$^{1}$, D. Dornic$^{1}$, F. Jouvenot$^{2}$, G. Maurin$^{3}$, for the KM3NeT consortium$^4$}
\afiliations{$^1$CPPM - Centre de Physiques des Particules de Marseille, CNRS/IN2P3 et Université de Mediterranée,
 163 avenue de Luminy, Case 902, 13288 Marseille Cedex 9,France\\ $^2$ University of Liverpool, Oliver Lodge 
Laboratory, L69 7ZE, Liverpool, United Kingdom\\ $^3$ DSM/DAPNIA - Direction des Sciences de la Matiere, 
Laboratoire de Recherche sur les lois fondamentales de l'Univers, CEA Saclay, 91191 Gif-Sur-Yvette Cedex, 
France\\ $^4$ This study was supported by the European Commission through the KM3NeT Design Study, FP6 contract no. 011937.}
\abstract{
Theoretical predictions for neutrino fluxes indicate that km$^{3}$ scale detectors are needed to detect certain 
astrophysical sources. The three Mediterranean experiments, ANTARES, NEMO and NESTOR are working together on a 
design study, KM3NeT, for a large deep-sea neutrino telescope. A detector placed in the Mediterranean Sea will 
survey a large part of the Galactic disc, including the Galactic Centre. It will complement the IceCube 
telescope currently under construction at the South Pole. Furthermore, the improved optical properties of sea water, compared to Antarctic ice, will allow a better angular resolution and hence better background rejection.

The main work presented in this paper is to evaluate different km$^{3}$ scale detector geometries in order to 
optimize the muon neutrino sensitivity between 1 and 100 TeV. For this purpose, we have developed a detailed 
simulation based on the {\it Mathematica} software - for the muon track production, the light transmission in water, 
the environmental background and the detector response. To compare different geometries, we have mainly used 
the effective neutrino area obtained after the full standard reconstruction chain.}

\begin{document}
\maketitle

\section{Introduction}
\vspace*{-0.4cm}
Most astrophysical source models predict neutrino production rates in addition to those of photons and cosmic rays. 
The goal of neutrino astronomy is to discover and measure these high energy neutrinos. Neutrinos offer 
several advantages over traditional astronomical messengers. They have a weak interaction 
cross-section, are stable and neutral. So, they can propagate through the Universe nearly unaffected 
by the environment and, contrary to cosmic-rays, without being deflected by magnetic fields. 

The recent results from gamma astronomy (especially the H.E.S.S. telescopes) and theoretical predictions 
for neutrino fluxes indicate that at least km$^{3}$ scale detectors are needed to detect certain 
astrophysical sources. For this reason a European consortium including the three Mediterranean experiments, ANTARES, NEMO and NESTOR is working on a design study for a large deep-sea 
neutrino telescope: KM3NeT {\cite{KM3NET-1}}. This detector will be placed in the northern hemisphere (in the Mediteranean Sea) to survey a large part of the Galaxy, including the Galactic Centre. It will so be complementary with the IceCube telescope currently under construction at the South Pole.

This paper presents \textsc{NESSY}, a new fast and detailed semi-analytic simulation based on the {\it Mathematica} 
software\footnote{Wolfram Research Inc. - www.wolfram.com}. Its purpose is to evaluate and optimize different km$^{3}$ scale detector geometries for neutrino 
detection in the Mediterranean sea between 1\,TeV and 1\,PeV. The first part of this report describes the Monte-Carlo 
simulations of muon tracks inside the detector and the reconstruction algorithm used. In the second part, a hexagonal 
detector was used to determine the optimal effective area for muon neutrino between 1\,TeV 
and 1\,PeV as a function of the photomultiplier (PMT) density. 

\vspace*{-0.4cm}
\section{NESSY: a full simulation and analysis chain}
\vspace*{-0.4cm}
When searching for sources, two main parameters are important: the sensitivity is proportional to the square of 
the angular resolution, while the number of reconstructed events is reflected in the effective area of the 
detector. To determine the best geometry, we need to optimize these two parameters. A full Monte-Carlo simulation and analysis chain has been developed with {\it Mathematica}. 

This software provides a fast and flexible tool to easily scan a large parameter space, including design 
properties (distance between lines and storeys, PMTs configurations) and the environmental 
characteristics (water optical properties, bioluminescence background).

\vspace*{-0.4cm}
\subsection{MC generator}
\vspace*{-0.3cm}
The simulation is organized in three different steps. The first step is the generation of muon tracks in a volume 
3 times the attenuation length (~30\,m) of the Cherenkov light in water around the instrumented volume. The second step 
is the propagation of the light produced by the muon and secondary particles induced by electromagnetic showers 
in the sea water. The light scattering in the medium is simulated with an analytic model of diffusion in sea water. Photons arriving on a PMT are converted into hits 
according to PMT properties (transit time spread, quantum efficiency and the electronics response).
Finally, the last step of the MC generation is to add the environmental background (K40 beta decay and bioluminescence). 

Several geometries may be considered for the future KM3NeT detector
including homogeneous or alternative
versions such as the ring or clustered configuration. Each configuration is constituted by vertical lines composed of a number of storeys, each containing a number of PMTs with given orientation and characteristics. 

\vspace*{-0.4cm}
\subsection{Selection, reconstruction \& analysis}
\vspace*{-0.3cm}
The first step is a rough selection of the hits. Only those in clusters of a minimum of 3 hits in adjacent storeys 
are considered. Furthermore, a causality filter relative to the barycentre of selected hits is used. The 
different steps in the hits selection procedure results in a purity better than 99\% of true hits.

The objective of the reconstruction is to determine the muon track characteristics - the zenith and azimuth angles - using the recorded arrival time and the number of hits. Due to the non linear dependence of the measurement, an iterative method is used to determine the best fit of track parameters. The reconstruction algorithm contains different consecutive fitting procedures using the result of the previous one as the starting point. The last fit uses a maximum likelihood method for which the probability density function was built with the time residual histogram obtained from MC data. 

After the reconstruction, the muon effective area is calculated from the
product of the geometric surface and the ratio of the number of reconstructed to generated events. To convert the effective area for muons to neutrinos, the following formula is used.
\begin{eqnarray}\label{equ1}
\ A_\nu(\theta,E_\nu)=A_\mu(\theta,E_\mu).R_\mu(E_\mu)\times N_A.\sigma_\nu(E_\nu)  \nonumber \\
  \times P_{\oplus}(\theta,E_\nu).I(E_\mu,E_\nu)~~~~~~~~ \nonumber
\end{eqnarray}
where $A_\nu(\theta,E_\nu)$ and $A_\mu(\theta,E_\mu)$ are the effective area respectively for neutrinos and muons, $R_\mu(E_\mu)$ is the effective muon range, $N_A$ is the Avogadro number, $\sigma_\nu(E_\nu)$ is the interaction cross section, $P_{\oplus}(\theta,E_\nu)$ is the absorption probability in the Earth and $I(E_\mu,E_\nu)$ is the conversion between neutrino and muon energy.

\begin{figure}
\begin{center}
\includegraphics [width=0.45\textwidth]{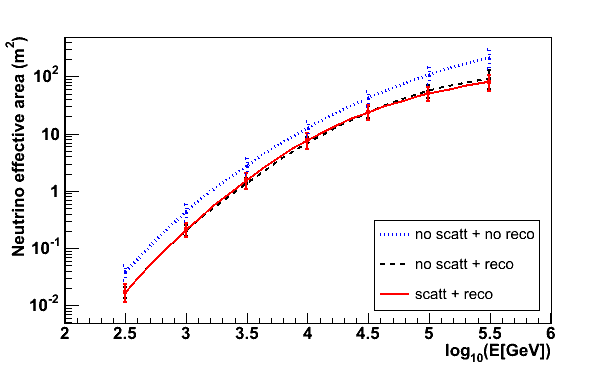}
\end{center}
\vspace*{-0.5cm}\caption{Neutrino effective area calculated for a hexagonal km$^{3}$ detector with different conditions: without scattering and before the reconstruction (blue), without scattering after the reconstruction (black) and with the scattering after the reconstruction (red).}\label{fig1}
\end{figure}

The figure \ref{fig1} presents the neutrino effective area calculated at
2 different steps: after the selection and after the reconstruction.
This study was made using a hexagonal KM3 detector composed of 127 lines
spaced by 100\,m (25 storeys each 15\,m). The reconstruction efficiency
increases with the energy and its mean value is around 50\%. This figure
shows also the influence of the delayed scattered hits for the
reconstruction processes on the neutrino effective area. Including the
scattered hits does not change the effective surface. Also, this doesn't
have a significant effect on the detector angular resolution (mean
difference of the angle between the reconstructed and the true muon
track) : 0.08 to 0.09\,$^\circ$ around 5 TeV. So, to reduce CPU time,
the scattering mode was disabled for the following studies.

\vspace*{-0.4cm}
\section{Homogeneous and compact geometry optimisation}
\vspace*{-0.4cm}
Based on ANTARES technologies, a homogeneous and almost isotropic detector has been optimized for neutrino detection
capability above 1\,TeV. The chosen configuration for this study is a homogeneous hexagon with 127 lines, each with 25 storeys with ANTARES type optical modules (3 $\times$ 10" PMTs \cite{OM}). 
For this work, the environmental parameters of the ANTARES site were used. In particular, an absorption length of 30 m (integrated over the Cherenkov light and the PMT spectral response) and a 40K background of 100\,KHz have been considered.

To obtain the best neutrino effective area, we have adopted a hierarchical procedure: first finding the optimal distance between lines, then, using this distance to find the optimal distance between storeys then finally finding the best PMT configuration in the storey. In order to have a fair comparison between different detectors, cuts in the selection procedure were ajusted for each tested geometry. For the two first steps, the number of lines and PMTs in the hexagonal detector remains constant. 

\vspace*{-0.4cm}
\subsection{The distance between lines}
\begin{figure}
\begin{center}
\includegraphics [width=0.45\textwidth]{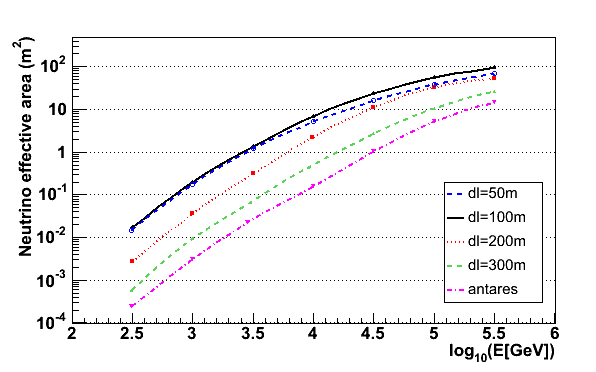}
\end{center}
\vspace*{-0.5cm}\caption{Effect of the distance between lines on the neutrino effective area (here, the number of lines and of PMTs remains constant).}\label{fig2}
\end{figure}

\vspace*{-0.3cm}
Figure \ref{fig2} shows the neutrino effective area calculated for different distances between lines (25 storeys separated by 15\,m). The effective area is a compromise between the geometrical surface which grows with the distance between lines and the detection efficiency which falls with this distance. Indeed, the most compact geometry presents the best result below a few TeV, but at higher energy, the configuration with 100\,m between lines has the optimal surface. The angular resolution and the detection efficiency decrease with the energy. The mean angular resolution of the detector at around 5\,TeV is 0.06, 0.09, 0.21 and 0.40\,$^\circ$ respectively for the configuration with 50, 100, 200 and 300\,m spacing.

\vspace*{-0.4cm}
\subsection{The distance between storeys}
\begin{figure}
\begin{center}
\includegraphics [width=0.45\textwidth]{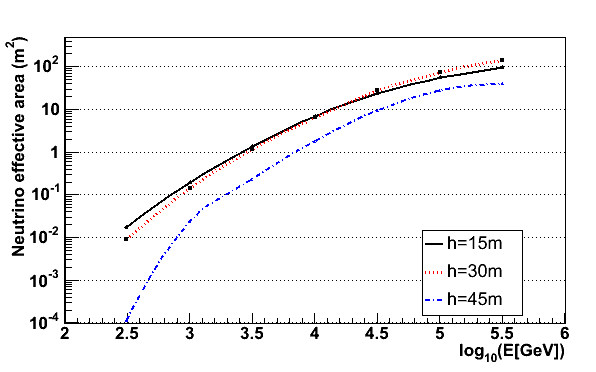}
\end{center}
\vspace*{-0.5cm}\caption{Effect of the distance between storeys on the neutrino
effective area (the PMT count remains constant in this study, 
only the height of the line is varying).}\label{fig3}
\end{figure}

\vspace*{-0.3cm}
Figure \ref{fig3} shows the effect of the distance between storeys on the effective area. The optimal distance is
15\,m for energy neutrino detection below 10\,TeV and 30\,m for higher
energy. The angular resolution remains constant for these two distances.
For larger distances, the selection
procedure becomes inefficient. In particular, the time windows to make clusters of hits become too large for background rejection. The neutrino flux of astrophysical sources decreases as a power law with the energy, so the configuration with 100\,m between lines and 15\,m between storeys is found as optimal. 

\vspace*{-0.4cm}
\subsection{The number of PMTs per storey}
\begin{figure}
\begin{center}
\includegraphics [width=0.45\textwidth]{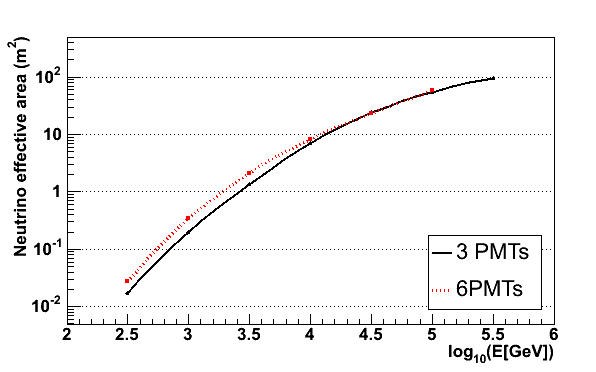}
\end{center}
\vspace*{-0.5cm}\caption{Effect of the PMT count in the storey on the neutrino effective area (the lines and storeys numbers remain constant).}\label{fig4}
\end{figure}

\vspace*{-0.3cm}
Figure \ref{fig4} shows the effective area for the optimal configuration with 3 and 6 optical modules in a storey. Doubling the PMT detection surface increases the neutrino effective area at low energy only by a factor around 1.6 (1\,TeV), but does not change it at high energy. As the neutrino spectrum of astrophysical sources decreases as a power law with the energy, the configuration with 6 optical modules is better, but more expensive.

\vspace*{-0.5cm}
\section{Summary and Discussion}
\begin{figure}
\begin{center}
\includegraphics [width=0.45\textwidth]{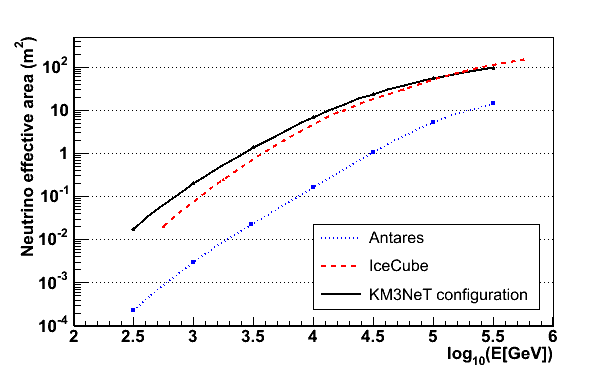}
\end{center}
\vspace*{-0.5cm}\caption{Preliminary neutrino effective area optimized with NESSY for the KM3NeT detector compared to Antares and IceCube {\cite{IceCube-1}} effective Area.}\label{fig5}
\end{figure}

\vspace*{-0.4cm}
Different configurations with a homogenous and almost isotropic detector have been tested to determine which one gives the optimal result for neutrino detection around 1 to 10\,TeV. A hexagonal geometry with 100\,m spaced lines and 3 PMTs per storey every 15\,m has the best effective area and an angular resolution better than 0.1$^\circ$. This result is compared to the Ice-Cube effective area {\cite{IceCube-1}} and is slightly better up to 10\,TeV (cf. figure \ref{fig5}). The ANTARES performance is also shown in reference.
This study is the first step to optimize the performance of a km$^{3}$
detector. Other parameters have to be taken into account such
as the environmental properties of the different sites, the effect of the energy reconstruction, etc. In the KM3NeT consortium,
others groups are testing some non homogenous geometries and have obtained comparable results\cite{Seb}. Moreover its performances have to be adjusted to enhance the detection of astrophysical sources {\cite{Carr}}.

\vspace*{-0.4cm}
%This is the reference to .bib file (Whitout .bib!)
\bibliographystyle{unsrt}
\bibliography{icrc0859}
%%%%%%%%%%%%%%%%%%%%%%%%%%%%%%%%%%%%%%%%%%%
%%  References
%%%%%%%%%%%%%%%%%%%%%%%%%%%%%%%%%%%%%%%%%%%

%This in the bibtex style, is ok.
\bibliographystyle{plain}

\end{document}